\documentclass[jkps,preprint,fleqn,showpacs,showkeys]{revtex4}
\usepackage{graphicx}
\usepackage{amssymb}
\usepackage{amsmath}
\usepackage{bm}
\newcommand{\dbar}{{\mathchar'26\mkern-11mu\mathrm{d}}}

\begin{document}
\setcounter{page}{0}
\title{Quantum Isothermal Reversible Process of Particles in a Box with a Delta
Potential}
\author{Minho \surname{Park}}
\author{Su Do \surname{Yi}}
\author{Seung Ki \surname{Baek}}
\email{seungki@pknu.ac.kr}
\thanks{Fax: +82-51-629-5549}
\affiliation{Department of Physics, Pukyong National University, Busan 608-737,
Korea}

\date[]{Received 25 November 2014}

\begin{abstract}
For an understanding of a heat engine working in the microscopic scale,
it is often necessary to estimate the amount of reversible work extracted by
isothermal expansion of the
quantum gas used as its working substance.
We consider an engine with a movable wall,
modeled as an infinite square well with a delta peak inside. By solving the
resulting one-dimensional Schr\"odinger equation,
we obtain the energy levels and the thermodynamic potentials.
Our result shows how quantum tunneling degrades the
engine by decreasing the amount of reversible work during the isothermal
expansion.
\end{abstract}
\pacs{03.65.Ge,05.70.Ce,07.20.Pe}
\keywords{Schr\"odinger equation, Quantum thermodynamics, Isothermal expansion}
\maketitle

\section{Introduction}

A fundamental idea behind classical statistical mechanics is that we are
living in a macroscopic world, which means that we are unable to deal
with microstates on a molecular level.
This forces us to distinguish heat from work and
to accept the increase of entropy on probabilistic ground, when we formulate
the laws of thermodynamics. However, the border between our macroscopic world
and the microscopic world of molecules is becoming vague due to
technological developments over the last century. This has led to a natural
question; i.e., what would it mean to thermodynamics if it became possible
to access and manipulate microstates just as we do macrostates? The founders of
statistical mechanics were already aware of this problem: For example,
Maxwell imagined an intelligent being's intervention on a molecular level
in his famous thought experiment, now known as Maxwell's
demon~\cite{review}, and expressed deep concern about the foundation
of the second law of thermodynamics.

The Szilard engine has been regarded as the simplest implementation of
Maxwell's demon with a single-particle gas~\cite{sagawa}.
Its cycle consists of four processes; inserting a wall into the center of
the box, measuring the position of the particle, expanding the quantum gas
isothermally, and finally removing the wall.
Considering its microscopic nature, it is tempting to translate the cycle
into quantum-mechanical language as has been done
in Refs.~\onlinecite{bender} and \onlinecite{kieu} for other engines.
In fact, there are more reasons than that: Zurek~\cite{zurek},
when he discusses Jauch and Baron's objection that the gas
is compressed without the expenditure of energy upon inserting the wall,
argues that one has to resort to quantum mechanics to understand the Szilard
engine~\cite{jauch}.
According to Ref.~\onlinecite{zurek}, this `apparent
inconsistency' is removed by quantum-mechanical considerations.
Therefore, in a sense, it is a matter of theoretical consistency.
Kim and coworkers have presented a detailed account for its quantum-mechanical
cycle in Refs.~\onlinecite{swkim,kim2,khkim},
with emphasis on the isothermal process.
There seem to remain some subtle issues to settle,
however, as shown in the debate on how to deal with the quantum
tunneling effect through the wall in a three-boson case~\cite{comment}.

In this work, we describe the isothermal expansion of the quantum Szilard
engine by considering a quantum gas confined
in a one-dimensional cylinder~\cite{vugalter,pedram}.
Although some previous studies, such as Refs.~\onlinecite{bender2,dong,li},
assume that the wall is impenetrable, we will relax that assumption because
it is, strictly speaking, experimentally infeasible.
Starting from the Schr\"odinger equation, we
calculate the energy levels and thereby obtain how much work can be extracted by
an isothermal reversible expansion. We find the following: The amount of
reversible work is significantly reduced compared to the claims in
Refs.~\onlinecite{zurek} and \onlinecite{swkim}
if we assume full thermal equilibrium inside the box,
as is usual in describing isothermal expansion.
This conclusion should not be affected by either
the wall's insertion or the measurement before the isothermal expansion
because thermal equilibrium does not remember its history.
Moreover, if we put two or three fermions in the box, the free-energy landscape
is not monotonic with respect to the wall's position,
so that we should sometimes \emph{perform} work to expand the gas.
These effects have not been explicitly discussed in previous studies.

This work is organized as follows: An explanation of our basic setting in terms
of the Schr\"odinger equation is given in Section~\ref{sec:isoexp}.
A numeric calculation of the energy levels and the free energy
and a comparison of a single-particle case to the two- and three-particle cases
is presented in in Section~\ref{sec:work}. This is followed by
discussion of results and conclusions.

\section{Particle in a box}
\label{sec:isoexp}

\begin{figure}
\includegraphics[width=0.45\textwidth]{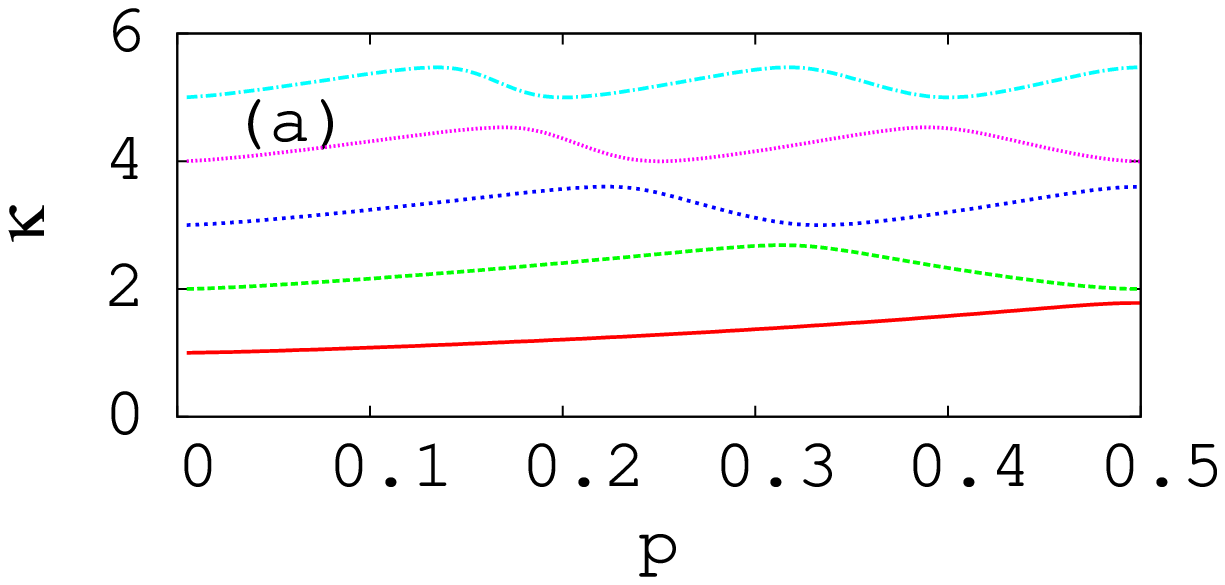}
\includegraphics[width=0.45\textwidth]{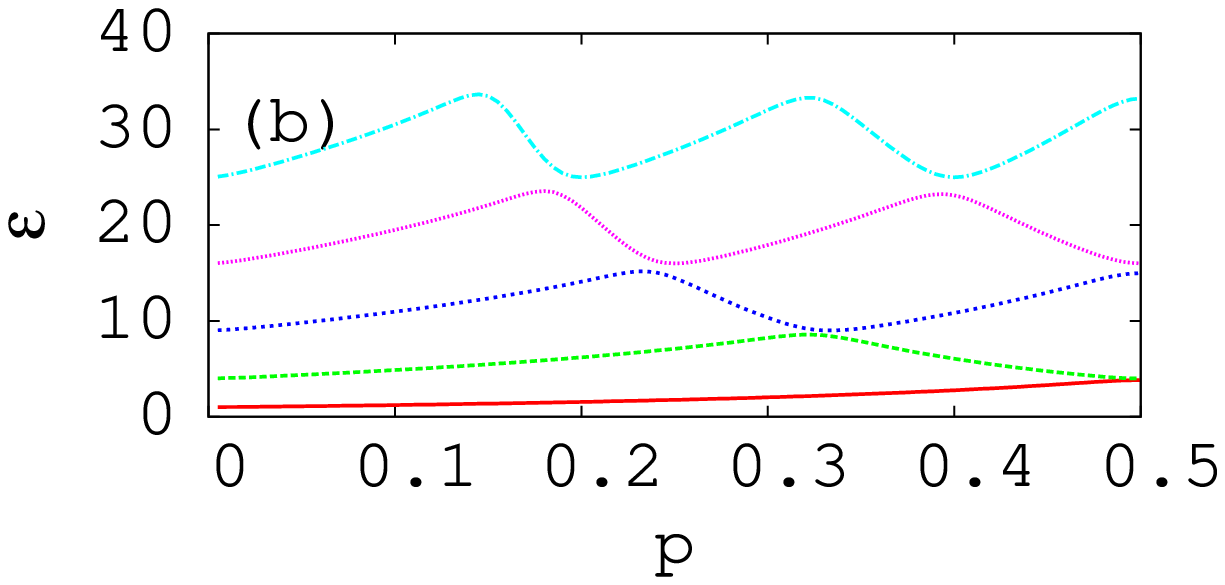}
\includegraphics[width=0.45\textwidth]{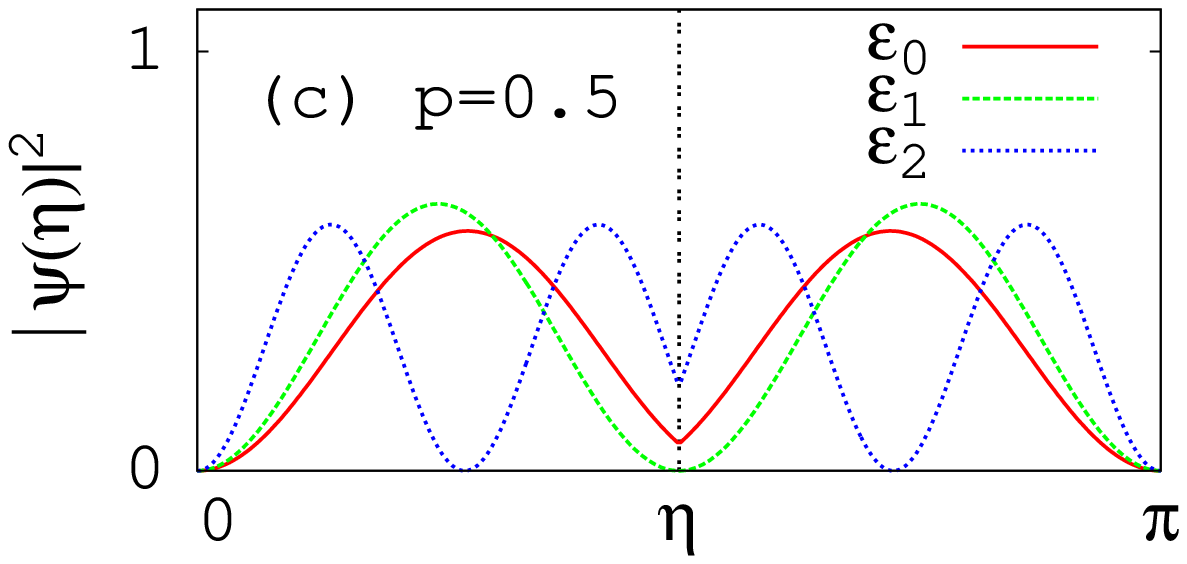}
\includegraphics[width=0.45\textwidth]{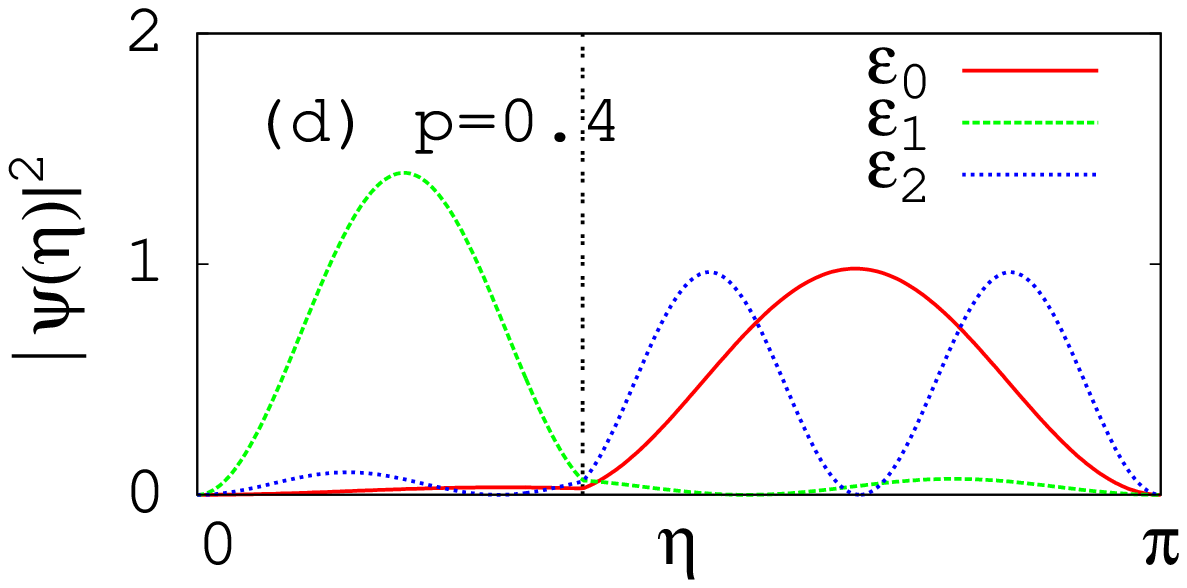}
\caption{(Color online) (a) Lowest wavenumbers in units of $\pi/L$ as a
function of $p$, and (b) the corresponding energy
levels in units of $E^{(0)}_g = \hbar^2 \pi^2 / (2mL^2)$.
The other two panels show the
probability densities for the three lowest eigenmodes
(c) when the wall is in the middle, $p=0.5$,
and (d) when it has moved to $p=0.4$.
The strength of the potential is set as $\alpha = 10$.}
\label{fig:level5}
\end{figure}

Consider the following one-dimensional potential landscape with size $L$:
\begin{equation}
V(x) = \left\{
\begin{array}{lc}
a \delta(x-pL) & \mbox{if~}0<x<L\\
\infty & \mbox{elsewhere},
\end{array}
\right.
\end{equation}
where $a$ is the strength of the delta potential and
$p \in (0,1)$ specifies its location~\cite{griffiths}.
We expect the Schr\"odinger equation
\begin{equation}
\left[-\frac{\hbar^2}{2m} \frac{d^2}{dx^2} + V(x) \right] \psi(x) = E \psi(x)
\label{eq:schrod}
\end{equation}
to have a solution with a sinusoidal form.
The boundary conditions at $x=0$ and $x=L$ are satisfied if we set
\begin{equation}
\psi(x) = \left\{
\begin{array}{lc}
A \sin kx & \mbox{if~} 0<x<pL\\
B \sin k(L-x) & \mbox{if~} pL<x<L,\\
\end{array}\right.
\label{eq:psi}
\end{equation}
where the coefficients $A$ and $B$, as well as the wavenumber $k$, are assumed
to be nonzero.
The eigenfunction $\psi(x)$ has the following properties: First, it is
continuous over $[0, L]$; i.e.,
\begin{equation}
A \sin pkL = B \sin qkL
\label{eq:cont}
\end{equation}
with $q \equiv 1-p$.
Second, the change in the derivative of $\psi(x)$ with respect to $x$
around $x=pL$ is obtained by using
\begin{equation}
\lim_{\epsilon \rightarrow 0} \left\{ \int_{pL-\epsilon}^{pL+\epsilon} \left[
\frac{d^2}{dx^2} \psi(x) \right] dx + \frac{2m}{\hbar^2} \left[
\int_{pL-\epsilon}^{pL+\epsilon} E \psi(x) dx - \int_{pL-\epsilon}^{pL+\epsilon}
a \psi(x) \delta(x-pL) dx \right] \right\} = 0,
\end{equation}
from which it follows that
\begin{equation}
\lim_{\epsilon \rightarrow 0}
\left(
\left. \frac{d\psi}{dx} \right|_{pL+\epsilon} -
\left. \frac{d\psi}{dx} \right|_{pL-\epsilon}
\right) =
\frac{2ma}{\hbar^2} \psi(pL).
\end{equation}
Plugging Eq.~(\ref{eq:psi}) here
and multiplying both sides by $\sin qkL$ to use Eq.~(\ref{eq:cont}), we obtain
\begin{equation}
- k \sin kL = \frac{2ma}{\hbar^2} \sin pkL \sin qkL.
\label{eq:k}
\end{equation}
This formula has also been derived in Ref.~\onlinecite{pedram} in a different
way.
Equation~(\ref{eq:k}) can be numerically solved by using the Newton-Raphson
method to yield the allowed values of $k$. The method works as follows:
First, let us define
$f(k) \equiv k \sin kL + \frac{2ma}{\hbar^2} \sin pkL \sin qkL$,
and look for its zeros to solve Eq.~(\ref{eq:k}). Starting from
$k=0$, we check whether the sign of $f(k)$ changes with increasing $k$ by a
sufficiently small amount, say, $dk \ll 1$. Every time the sign changes
between $k=k^\ast > 0$ and $k^\ast +dk$,
we run the following iteration starting from $k^{(0)} = k^\ast$:
\begin{equation}
k^{(j+1)} = k^{(j)} - \frac{f[k^{(j)}]}{f'[k^{(j)}]},
\end{equation}
where $j=0, 1, 2, \ldots$. When $k^{(j)}$ has converged to a stationary value
$k^{(\infty)}$,
i.e., if $k^{(j)} = k^{(j+1)}$ within numerical accuracy, we take it as a
solution of Eq.~(\ref{eq:k}) and repeat the procedure by checking the sign of
$f(k)$ from $k=k^{(\infty)}+dk$. Each of these solutions is identified
with the wavenumber $k_n$ of the $n$th eigenmode in an ascending order.
For the application of the Newton-Raphson method numerically,
it is convenient to
express the quantities in dimensionless units, so we divide
Eq.~(\ref{eq:schrod}) by $E^{(0)}_g \equiv \hbar^2 \pi^2 / (2mL^2)$, the
ground state energy for $a=0$, to obtain
\begin{equation}
\left[ \frac{d}{d\eta^2} \psi(\eta) + \Phi(\eta) \right] \psi(\eta) = \epsilon
\psi(\eta),
\end{equation}
where $\eta \equiv \pi x / L$. Then, the potential inside the box
is expressed as
\begin{equation}
\Phi(\eta) = \frac{2mL^2}{\hbar^2 \pi^2} a \delta(x-pL)
= \alpha \delta(\eta - p\pi),
\end{equation}
with $\alpha \equiv 2a mL / (\hbar^2 \pi) = a (L/\pi)
/ E^{(0)}_g$. We can also define a dimensionless
wavenumber $\kappa \equiv L k / \pi$ to rewrite Eqs.~(\ref{eq:k}) as
$\kappa \sin \kappa \pi + \alpha \sin p \kappa \pi \sin q \kappa \pi = 0$.
A little algebra shows that the energy level is given by
\begin{equation}
\epsilon = \kappa^2 + \frac{2\alpha}{\pi} \left[ \csc^2 p\kappa \pi
\left(p - \frac{1}{2\kappa \pi} \sin 2p \kappa \pi \right) + \csc^2
q \kappa \pi \left( q - \frac{1}{2 \kappa \pi}
\sin 2q \kappa \pi \right) \right]^{-1},
\end{equation}
where the second term represents the potential energy coming from the
overlap between the wavefunction and the delta peak potential.
The numerical solutions for these $\kappa$ and
$\epsilon$ are depicted in Fig.~\ref{fig:level5}, together with
the probability density plots.

\section{Reversible work}
\label{sec:work}

\subsection{Single Particle}

We are interested in how much work can be extracted by performing a
reversible isothermal process with this system.
Let $\epsilon_n$ denote the $n$th energy level obtained by using
the Newton-Raphson method with $n=1,\ldots,M$ so that the ground-state energy
is denoted by $\epsilon_1$.
For a given value of temperature $T = (k_B \beta)^{-1}$,
where $k_B$ is the Boltzmann
constant, we run the calculation to get $\beta \epsilon_M \gg 1$.
The partition function for a single quantum particle can then be obtained as
\begin{equation}
Z = \lim_{M \rightarrow \infty} \sum_{n=1}^M e^{-\beta \epsilon_n}.
\label{eq:z}
\end{equation}
The free energy is
$F = -k_B T \ln Z$,
accompanied by a differential form
$dF = -S dT - P dV$,
where $S$, $P$, and $V$ are the entropy, pressure, and volume, respectively.
The classical concepts, such as pressure and work, need careful consideration
in quantum mechanics~\cite{pressure}. To calculate work,
we begin with the occupation probability of the $n$th level given by
\begin{equation}
\rho_n = \frac{e^{-\beta \epsilon_n}}{\sum_n e^{-\beta \epsilon_n}} =
\frac{-k_B T \left( \partial Z / \partial \epsilon_n \right)}{Z}
= \frac{\partial}{\partial \epsilon_n} (-k_B T \ln Z) =
\frac{\partial F}{\partial \epsilon_n}.
\end{equation}
The change in the internal energy $U = \sum_n \rho_n \epsilon_n$ is then
expressed as
$dU = \sum_n \epsilon_n d\rho_n + \rho_n d\epsilon_n = \dbar Q + \dbar W$,
where $\dbar Q$ and $\dbar W$ are differentials of heat and work, respectively.
The quantum thermodynamic work is identified with
$\dbar W = \sum_n \rho_n d\epsilon_n$
because heat is usually involved with changes in occupation probabilities
for given energy levels whereas work is performed when the energy
levels themselves change~\cite{swkim} .
Consequently, the amount of reversible work during this process is
$W = \int \sum_n \frac{\partial F}{\partial \epsilon_n} d\epsilon_n = \int dF =
\Delta F$,
which is consistent with the classical case.

\begin{figure}
\includegraphics[width=0.45\textwidth]{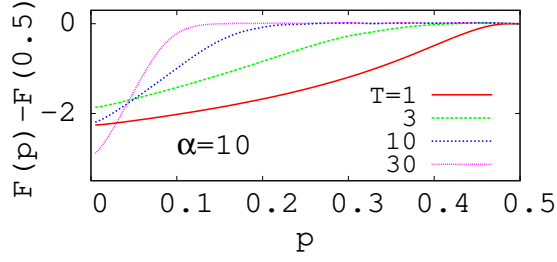}
\caption{(Color online)
Change in the free energy $F$ in units of $E^{(0)}_g = \hbar^2 \pi^2 / (2mL^2)$
when the wall, starting from the center of the box, moves to a new position $p$.
The height of the wall is fixed at $\alpha=10$, and each
curve represents a different $T$ in units of $E^{(0)}_g / k_B$.
The number of energy levels considered is kept as
$M=10^3$ throughout the free-energy calculations.}
\label{fig:df}
\end{figure}

Figure~\ref{fig:df} shows $\Delta F$ resulting from moving the wall, which was
initially located at the center, $p=0.5$.
Suppose that $T$ is so low that $T \lesssim E^{(0)}_g/k_B$. When the wall is
at $p=0$, only the ground state contributes to the summation in
Eq.~(\ref{eq:z}), which means that $Z(p=0) \approx e^{-\beta E^{(0)}_g}$.
When the wall is at $p=0.5$, the ground-state energy equals $4 E^{(0)}_g$,
and the first excited state also contributes to the partition sum
in Eq.~(\ref{eq:z}) because it lies close to the ground state
[see Fig.~\ref{fig:level5}(b)]. The partition function
is, therefore, approximated as $Z(p=0.5) \approx 2 e^{-4\beta E^{(0)}_g}$.
The free-energy difference in this low-$T$ region is, thus, given as
$F(0)-F(0.5) \approx -3E^{(0)}_g + k_B T \ln 2$. When $T = E^{(0)}_g/k_B$,
for example, the free-energy difference
roughly amounts to $-2.3 E^{(0)}_g$, which explains the size
of the free-energy drop in Fig.~\ref{fig:df}.

As $T$ increases, a plateau develops near the center of the box
because $F$ does not respond much to the wall's position.
However, the drop in $F$ at the end of the process increases with increasing
$T$ if $T \gg O(1)$ in units of $E^{(0)}_g/k_B$.
Our question is how it grows with $T$.
Let us take $\alpha \rightarrow \infty$ and $\gamma \equiv \beta
\hbar^2 \pi^2 / (2mL^2) \ll 1$. Then, it is the
partition function for the initial and final wall positions can be
approximated. The latter case
of the final wall position is estimated as
\begin{eqnarray}
Z(p=0) &=& \sum_{n=1}^\infty e^{-\gamma n^2}\\
&\approx& \int_0^{\infty} \exp(-\gamma n^2) dn - \frac{1}{2} \sum_{n=0}^\infty
\left[ e^{-\gamma n^2} - e^{-\gamma (n+1)^2} \right]\label{eq:correction}\\
&=& \sqrt{\frac{\pi}{4\gamma}} - \frac{1}{2},
\end{eqnarray}
where the summation in Eq.~(\ref{eq:correction}) is a first-order
correction to the integral approximation.
Likewise, the former case of the initial wall position gives
\begin{equation}
Z(p=0.5) = 2 \sum_{n=1}^\infty e^{-4\gamma n^2}
\approx 2 \left[ \int_0^{\infty} \exp(-4\gamma n^2) dn - \frac{1}{2} \right]
= \sqrt{\frac{\pi}{4\gamma}} - 1,
\end{equation}
where the factor of $2$ in front of the summation is due to the fact that
every pair of adjacent energy levels becomes degenerate
in the limit of $\alpha \rightarrow \infty$.
This effect cancels the factor of $4$ inside the exponential arising
from the reduced volume $L/2$.
Although $\Delta F$ does not completely vanish due to the correction,
the important point is that $Z(0.5)/Z(0)$ for $\gamma \ll 1$ is far less than
$2$, in contrast
with the claim that $\Delta F = -k_B T \ln 2$ in Refs.~\onlinecite{zurek}
and \onlinecite{swkim}.
Because $\alpha$ must be finite in any experimental situation,
the particle can, in full thermal equilibrium, be observed on either side of
the box, which reduces the amount of extracted work. On the other hand,
the previous studies on the quantum Szilard engine~\cite{zurek,swkim}
have assumed that the process is performed within a shorter time scale
than required for tunneling through the wall and, therefore, concluded
that $Z(p=0.5) \approx \int \exp(-4\gamma n^2) dn$ without the factor of $2$
in front.
Strictly speaking, their engine undergoes an isothermal process
only when partially equilibrated to maintain the confinement and extract
$W= k_B T \ln 2$.

\subsection{Two- and Three-particle Cases}

\begin{figure}
\includegraphics[width=0.45\textwidth]{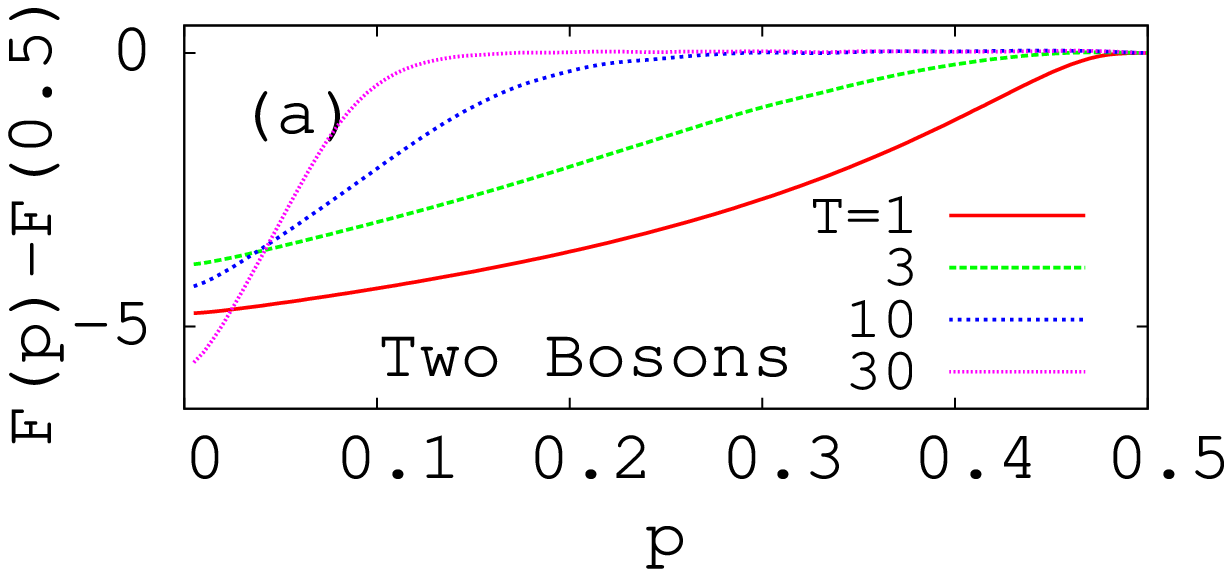}
\includegraphics[width=0.45\textwidth]{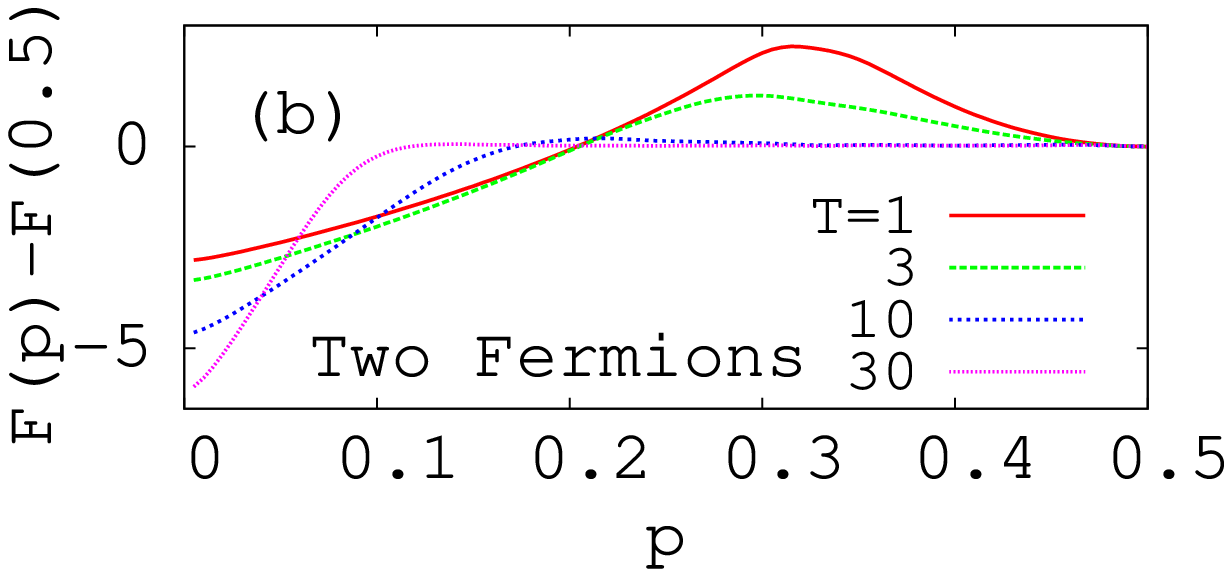}
\includegraphics[width=0.45\textwidth]{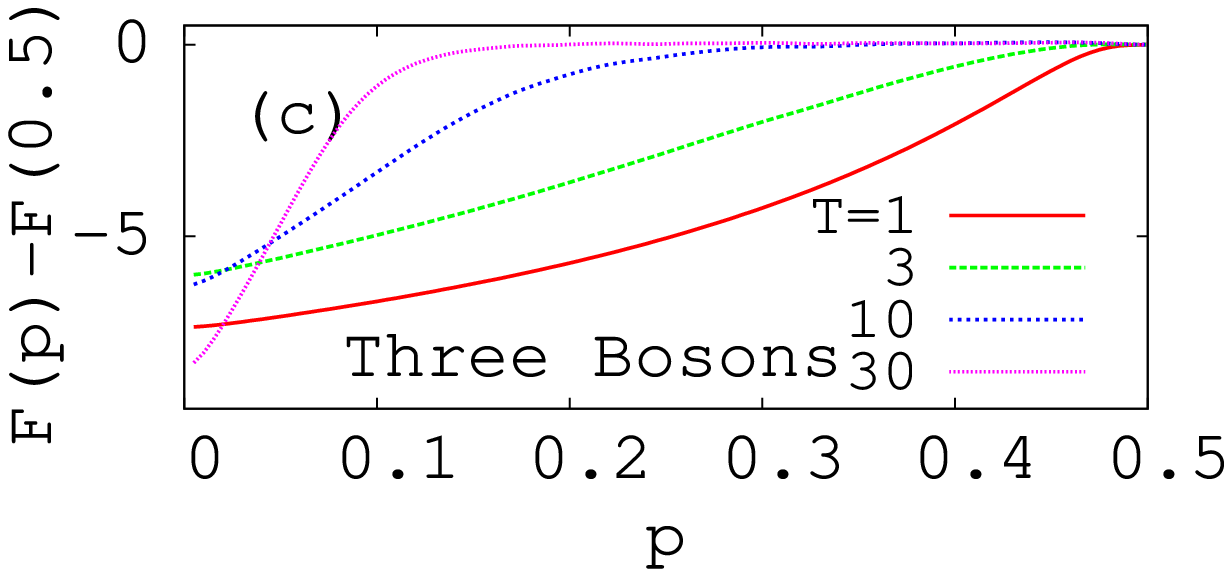}
\includegraphics[width=0.45\textwidth]{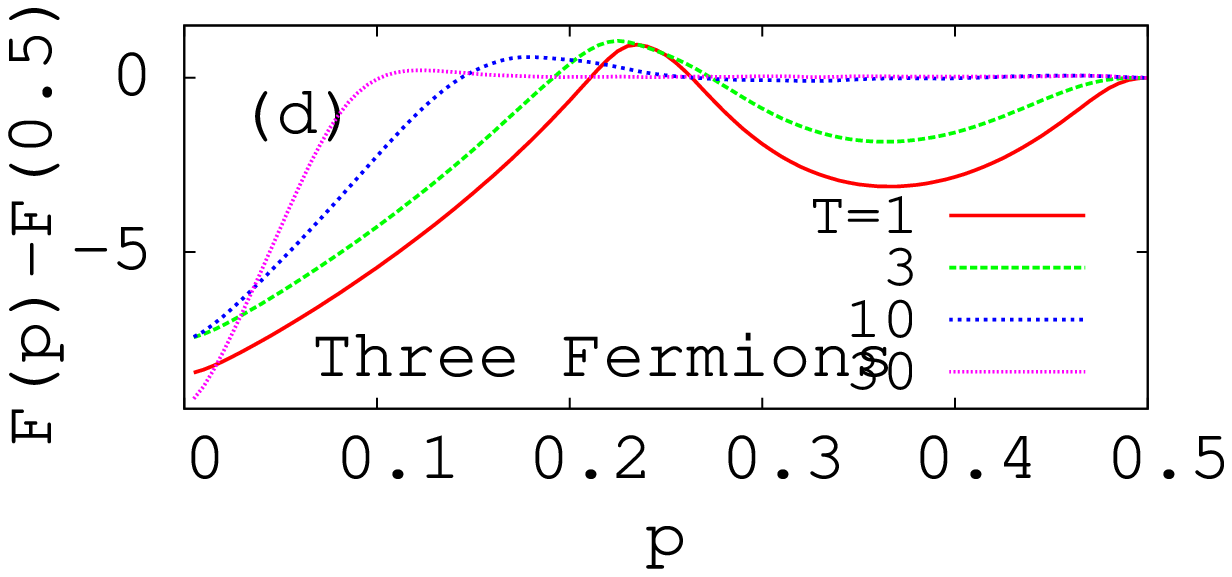}
\caption{(Color online) Free-energy changes in the presence of multiple
non-interacting quantum particles: (a) bosonic and (b) fermionic cases with two
particles. With three particles, (c) the bosonic case remains qualitatively
similar, but (d) the fermions behave differently due to the Pauli exclusion
principle.}
\label{fig:sym}
\end{figure}

If two quantum particles are in the box, we should consider their
symmetry, i.e., whether they are fermions or bosons.
Without any consideration of the spin degree of freedom,
the elements of the density matrix for the two particles are expressed
in the bracket notation as follows:
$\left< \epsilon_1' \epsilon_2' \left| e^{-\beta\hat{H}} \right|
\epsilon_1 \epsilon_2 \right> = e^{ -\beta (\epsilon_1 + \epsilon_2) }
\left( \delta_{\epsilon_1' \epsilon_1} \delta_{\epsilon_2' \epsilon_2} \pm
\delta_{\epsilon_1' \epsilon_2}
\delta_{\epsilon_2' \epsilon_1} \right)$,
where $\delta$ is the Kronecker delta symbol, and we have a plus sign for
bosons and a minus sign for fermions.
The two-particle partition function is then obtained by taking the trace
operation:
\begin{equation}
Z = \text{Tr}\left( e^{-\beta \hat{H}} \right)
= \frac{1}{2!} \sum_{\epsilon_1, \epsilon_2} e^{ - \beta
(\epsilon_1 + \epsilon_2) } \left( 1 \pm \delta_{\epsilon_1 \epsilon_2}
\right).
\end{equation}
The same applies to the three-particle case. The
density-matrix elements are written as
$\left< \epsilon_1' \epsilon_2' \epsilon_3' \left| e^{-\beta\hat{H}} \right|
\epsilon_1 \epsilon_2 \epsilon_3 \right> = e^{ -\beta (\epsilon_1 + \epsilon_2 +
\epsilon_3) }
\left( \delta_{\epsilon_1' \epsilon_1} \delta_{\epsilon_2' \epsilon_2}
\delta_{\epsilon_3' \epsilon_3} \pm \ldots \right)$,
and the partition function reads
\begin{equation}
Z = \text{Tr}\left( e^{-\beta \hat{H}} \right)
= \frac{1}{3!} \sum_{\epsilon_1, \epsilon_2, \epsilon_3} e^{ - \beta
(\epsilon_1 + \epsilon_2 + \epsilon_3) } \left( 1 \pm \delta_{\epsilon_1
\epsilon_2} \pm \delta_{\epsilon_2 \epsilon_3} \pm \delta_{\epsilon_3
\epsilon_1} + 2 \delta_{\epsilon_1 \epsilon_2} \delta_{\epsilon_2 \epsilon_3}
\right).
\end{equation}

Figure~\ref{fig:sym} shows the free-energy differences for the two- and the
three-particle cases. The bosonic cases are qualitatively similar to the
single-particle case. However, we see very different behavior for the fermionic
cases where the Pauli exclusion principle is in action: At low $T$, we need to
perform positive work to move the wall from the center to a certain position
$p$ [Figs.~\ref{fig:sym}(b) and (d)]. Obviously, the reason is that
it costs free energy to place two fermions close to each other.

\section{Summary and Outlook}

In summary, we have considered an isothermal expansion process of a quantum
gas, taking the tunneling effect into consideration.
We have found that the amount of reversible work is
smaller than $k_B T \ln 2$ in the high-$T$ region if the system is fully
equilibrated at every moment of the isothermal expansion. The difference from
Refs.~\onlinecite{zurek} and \onlinecite{swkim} arises because they have
separated the time scale of
tunneling from that of the partial equilibration that occurs on only one side
of the wall. Although it was not explicitly stated in previous studies, this
separation may be a plausible assumption for the following
reason: As we increase the potential height $\alpha$, we may well expect
the time scale for tunneling to grow whereas the partial equilibration before
tunneling is achieved within a finite amount of time. The quantum Szilard
engine will show its expected performance only between these two time scales.
The question is, then, how large a value of $\alpha$ one should have
to ensure the separation of the time scales,
which will be pursued in our future studies.

\acknowledgments
This work was supported by a research grant from Pukyong National University
(2014).


\begin{thebibliography}{17}%
\makeatletter
\providecommand \@ifxundefined [1]{%
 \@ifx{#1\undefined}
}%
\providecommand \@ifnum [1]{%
 \ifnum #1\expandafter \@firstoftwo
 \else \expandafter \@secondoftwo
 \fi
}%
\providecommand \@ifx [1]{%
 \ifx #1\expandafter \@firstoftwo
 \else \expandafter \@secondoftwo
 \fi
}%
\providecommand \natexlab [1]{#1}%
\providecommand \enquote  [1]{``#1''}%
\providecommand \bibnamefont  [1]{#1}%
\providecommand \bibfnamefont [1]{#1}%
\providecommand \citenamefont [1]{#1}%
\providecommand \href@noop [0]{\@secondoftwo}%
\providecommand \href [0]{\begingroup \@sanitize@url \@href}%
\providecommand \@href[1]{\@@startlink{#1}\@@href}%
\providecommand \@@href[1]{\endgroup#1\@@endlink}%
\providecommand \@sanitize@url [0]{\catcode `\\12\catcode `\$12\catcode
  `\&12\catcode `\#12\catcode `\^12\catcode `\_12\catcode `\%12\relax}%
\providecommand \@@startlink[1]{}%
\providecommand \@@endlink[0]{}%
\providecommand \url  [0]{\begingroup\@sanitize@url \@url }%
\providecommand \@url [1]{\endgroup\@href {#1}{\urlprefix }}%
\providecommand \urlprefix  [0]{URL }%
\providecommand \Eprint [0]{\href }%
\providecommand \doibase [0]{http://dx.doi.org/}%
\providecommand \selectlanguage [0]{\@gobble}%
\providecommand \bibinfo  [0]{\@secondoftwo}%
\providecommand \bibfield  [0]{\@secondoftwo}%
\providecommand \translation [1]{[#1]}%
\providecommand \BibitemOpen [0]{}%
\providecommand \bibitemStop [0]{}%
\providecommand \bibitemNoStop [0]{.\EOS\space}%
\providecommand \EOS [0]{\spacefactor3000\relax}%
\providecommand \BibitemShut  [1]{\csname bibitem#1\endcsname}%
\let\auto@bib@innerbib\@empty
\bibitem [{\citenamefont {Maruyama}\ \emph {et~al.}(2009)\citenamefont
  {Maruyama}, \citenamefont {Nori},\ and\ \citenamefont {Vedral}}]{review}%
  \BibitemOpen
  \bibfield  {author} {\bibinfo {author} {\bibfnamefont {K.}~\bibnamefont
  {Maruyama}}, \bibinfo {author} {\bibfnamefont {F.}~\bibnamefont {Nori}}, \
  and\ \bibinfo {author} {\bibfnamefont {V.}~\bibnamefont {Vedral}},\
  }\href@noop {} {\bibfield  {journal} {\bibinfo  {journal} {Rev. Mod. Phys.}\
  }\textbf {\bibinfo {volume} {81}},\ \bibinfo {pages} {1} (\bibinfo {year}
  {2009})}\BibitemShut {NoStop}%
\bibitem [{\citenamefont {Sagawa}\ and\ \citenamefont {Ueda}(2013)}]{sagawa}%
  \BibitemOpen
  \bibfield  {author} {\bibinfo {author} {\bibfnamefont {T.}~\bibnamefont
  {Sagawa}}\ and\ \bibinfo {author} {\bibfnamefont {M.}~\bibnamefont {Ueda}},\
  }in\ \href@noop {} {\emph {\bibinfo {booktitle} {Nonequilibrium Statistical
  Physics of Small Systems: Fluctuation Relations and Beyond}}},\ \bibinfo
  {editor} {edited by\ \bibinfo {editor} {\bibfnamefont {R.}~\bibnamefont
  {Klages}}, \bibinfo {editor} {\bibfnamefont {W.}~\bibnamefont {Just}}, \ and\
  \bibinfo {editor} {\bibfnamefont {C.}~\bibnamefont {Jarzynski}}}\ (\bibinfo
  {publisher} {Wiley},\ \bibinfo {address} {Weinheim},\ \bibinfo {year}
  {2013}),\ pp.\ \bibinfo {pages} {181--211}\BibitemShut {NoStop}%
\bibitem [{\citenamefont {Bender}\ \emph {et~al.}(2000)\citenamefont {Bender},
  \citenamefont {Brody},\ and\ \citenamefont {Meister}}]{bender}%
  \BibitemOpen
  \bibfield  {author} {\bibinfo {author} {\bibfnamefont {C.~M.}\ \bibnamefont
  {Bender}}, \bibinfo {author} {\bibfnamefont {D.~C.}\ \bibnamefont {Brody}}, \
  and\ \bibinfo {author} {\bibfnamefont {B.~K.}\ \bibnamefont {Meister}},\
  }\href@noop {} {\bibfield  {journal} {\bibinfo  {journal} {J. Phys. A}\
  }\textbf {\bibinfo {volume} {33}},\ \bibinfo {pages} {4427} (\bibinfo {year}
  {2000})}\BibitemShut {NoStop}%
\bibitem [{\citenamefont {Kieu}(2004)}]{kieu}%
  \BibitemOpen
  \bibfield  {author} {\bibinfo {author} {\bibfnamefont {T.~D.}\ \bibnamefont
  {Kieu}},\ }\href@noop {} {\bibfield  {journal} {\bibinfo  {journal} {Phys.
  Rev. Lett.}\ }\textbf {\bibinfo {volume} {93}},\ \bibinfo {pages} {140403}
  (\bibinfo {year} {2004})}\BibitemShut {NoStop}%
\bibitem [{\citenamefont {Zurek}(1986)}]{zurek}%
  \BibitemOpen
  \bibfield  {author} {\bibinfo {author} {\bibfnamefont {W.}~\bibnamefont
  {Zurek}},\ }in\ \href@noop {} {\emph {\bibinfo {booktitle} {Frontiers of
  Nonequilibrium Statistical Physics}}},\ \bibinfo {series} {NATO ASI Series},
  Vol.\ \bibinfo {volume} {135},\ \bibinfo {editor} {edited by\ \bibinfo
  {editor} {\bibfnamefont {G.~T.}\ \bibnamefont {Moore}}\ and\ \bibinfo
  {editor} {\bibfnamefont {M.~O.}\ \bibnamefont {Scully}}}\ (\bibinfo
  {publisher} {Plenum},\ \bibinfo {address} {New York},\ \bibinfo {year}
  {1986}),\ pp.\ \bibinfo {pages} {151--161}\BibitemShut {NoStop}%
\bibitem [{\citenamefont {Jauch}\ and\ \citenamefont {Baron}(1972)}]{jauch}%
  \BibitemOpen
  \bibfield  {author} {\bibinfo {author} {\bibfnamefont {J.~M.}\ \bibnamefont
  {Jauch}}\ and\ \bibinfo {author} {\bibfnamefont {J.~G.}\ \bibnamefont
  {Baron}},\ }\href@noop {} {\bibfield  {journal} {\bibinfo  {journal} {Helv.
  Phys. Acta}\ }\textbf {\bibinfo {volume} {45}},\ \bibinfo {pages} {220}
  (\bibinfo {year} {1972})}\BibitemShut {NoStop}%
\bibitem [{\citenamefont {Kim}\ \emph {et~al.}(2011)\citenamefont {Kim},
  \citenamefont {Sagawa}, \citenamefont {Liberto},\ and\ \citenamefont
  {Ueda}}]{swkim}%
  \BibitemOpen
  \bibfield  {author} {\bibinfo {author} {\bibfnamefont {S.~W.}\ \bibnamefont
  {Kim}}, \bibinfo {author} {\bibfnamefont {T.}~\bibnamefont {Sagawa}},
  \bibinfo {author} {\bibfnamefont {S.~D.}\ \bibnamefont {Liberto}}, \ and\
  \bibinfo {author} {\bibfnamefont {M.}~\bibnamefont {Ueda}},\ }\href@noop {}
  {\bibfield  {journal} {\bibinfo  {journal} {Phys. Rev. Lett.}\ }\textbf
  {\bibinfo {volume} {106}},\ \bibinfo {pages} {070401} (\bibinfo {year}
  {2011})}\BibitemShut {NoStop}%
\bibitem [{\citenamefont {Kim}\ and\ \citenamefont {Kim}(2011)}]{kim2}%
  \BibitemOpen
  \bibfield  {author} {\bibinfo {author} {\bibfnamefont {K.-H.}\ \bibnamefont
  {Kim}}\ and\ \bibinfo {author} {\bibfnamefont {S.~W.}\ \bibnamefont {Kim}},\
  }\href@noop {} {\bibfield  {journal} {\bibinfo  {journal} {Phys. Rev. E}\
  }\textbf {\bibinfo {volume} {84}},\ \bibinfo {pages} {012101} (\bibinfo
  {year} {2011})}\BibitemShut {NoStop}%
\bibitem [{\citenamefont {Kim}\ and\ \citenamefont {Kim}(2012)}]{khkim}%
  \BibitemOpen
  \bibfield  {author} {\bibinfo {author} {\bibfnamefont {K.-H.}\ \bibnamefont
  {Kim}}\ and\ \bibinfo {author} {\bibfnamefont {S.~W.}\ \bibnamefont {Kim}},\
  }\href@noop {} {\bibfield  {journal} {\bibinfo  {journal} {J. Korean Phys.
  Soc.}\ }\textbf {\bibinfo {volume} {61}},\ \bibinfo {pages} {1187} (\bibinfo
  {year} {2012})}\BibitemShut {NoStop}%
\bibitem [{\citenamefont {Plesch}\ \emph {et~al.}(2013)\citenamefont {Plesch},
  \citenamefont {Dahlsten}, \citenamefont {Goold},\ and\ \citenamefont
  {Vedral}}]{comment}%
  \BibitemOpen
  \bibfield  {author} {\bibinfo {author} {\bibfnamefont {M.}~\bibnamefont
  {Plesch}}, \bibinfo {author} {\bibfnamefont {O.}~\bibnamefont {Dahlsten}},
  \bibinfo {author} {\bibfnamefont {J.}~\bibnamefont {Goold}}, \ and\ \bibinfo
  {author} {\bibfnamefont {V.}~\bibnamefont {Vedral}},\ }\href@noop {}
  {\bibfield  {journal} {\bibinfo  {journal} {Phys. Rev. Lett.}\ }\textbf
  {\bibinfo {volume} {111}},\ \bibinfo {pages} {188901} (\bibinfo {year}
  {2013})}\BibitemShut {NoStop}%
\bibitem [{\citenamefont {Vugalter}\ \emph {et~al.}(2002)\citenamefont
  {Vugalter}, \citenamefont {Das},\ and\ \citenamefont {Sorokin}}]{vugalter}%
  \BibitemOpen
  \bibfield  {author} {\bibinfo {author} {\bibfnamefont {G.~A.}\ \bibnamefont
  {Vugalter}}, \bibinfo {author} {\bibfnamefont {A.~K.}\ \bibnamefont {Das}}, \
  and\ \bibinfo {author} {\bibfnamefont {V.~A.}\ \bibnamefont {Sorokin}},\
  }\href@noop {} {\bibfield  {journal} {\bibinfo  {journal} {Phys. Rev. A}\
  }\textbf {\bibinfo {volume} {66}},\ \bibinfo {pages} {012104} (\bibinfo
  {year} {2002})}\BibitemShut {NoStop}%
\bibitem [{\citenamefont {Pedram}\ and\ \citenamefont {Vahabi}(2010)}]{pedram}%
  \BibitemOpen
  \bibfield  {author} {\bibinfo {author} {\bibfnamefont {P.}~\bibnamefont
  {Pedram}}\ and\ \bibinfo {author} {\bibfnamefont {M.}~\bibnamefont
  {Vahabi}},\ }\href@noop {} {\bibfield  {journal} {\bibinfo  {journal} {Am. J.
  Phys.}\ }\textbf {\bibinfo {volume} {78}},\ \bibinfo {pages} {839} (\bibinfo
  {year} {2010})}\BibitemShut {NoStop}%
\bibitem [{\citenamefont {Bender}\ \emph {et~al.}(2005)\citenamefont {Bender},
  \citenamefont {Brody},\ and\ \citenamefont {Meister}}]{bender2}%
  \BibitemOpen
  \bibfield  {author} {\bibinfo {author} {\bibfnamefont {C.~M.}\ \bibnamefont
  {Bender}}, \bibinfo {author} {\bibfnamefont {D.~C.}\ \bibnamefont {Brody}}, \
  and\ \bibinfo {author} {\bibfnamefont {B.~K.}\ \bibnamefont {Meister}},\
  }\href@noop {} {\bibfield  {journal} {\bibinfo  {journal} {P. Roy. Soc. A}\
  }\textbf {\bibinfo {volume} {461}},\ \bibinfo {pages} {733} (\bibinfo {year}
  {2005})}\BibitemShut {NoStop}%
\bibitem [{\citenamefont {Dong}\ \emph {et~al.}(2011)\citenamefont {Dong},
  \citenamefont {Xu}, \citenamefont {Cai},\ and\ \citenamefont {Sun}}]{dong}%
  \BibitemOpen
  \bibfield  {author} {\bibinfo {author} {\bibfnamefont {H.}~\bibnamefont
  {Dong}}, \bibinfo {author} {\bibfnamefont {D.~Z.}\ \bibnamefont {Xu}},
  \bibinfo {author} {\bibfnamefont {C.~Y.}\ \bibnamefont {Cai}}, \ and\
  \bibinfo {author} {\bibfnamefont {C.~P.}\ \bibnamefont {Sun}},\ }\href@noop
  {} {\bibfield  {journal} {\bibinfo  {journal} {Phys. Rev. E}\ }\textbf
  {\bibinfo {volume} {83}},\ \bibinfo {pages} {061108} (\bibinfo {year}
  {2011})}\BibitemShut {NoStop}%
\bibitem [{\citenamefont {Li}\ \emph {et~al.}(2012)\citenamefont {Li},
  \citenamefont {Zou}, \citenamefont {Li}, \citenamefont {Shao},\ and\
  \citenamefont {Wu}}]{li}%
  \BibitemOpen
  \bibfield  {author} {\bibinfo {author} {\bibfnamefont {H.}~\bibnamefont
  {Li}}, \bibinfo {author} {\bibfnamefont {J.}~\bibnamefont {Zou}}, \bibinfo
  {author} {\bibfnamefont {J.-G.}\ \bibnamefont {Li}}, \bibinfo {author}
  {\bibfnamefont {B.}~\bibnamefont {Shao}}, \ and\ \bibinfo {author}
  {\bibfnamefont {L.-A.}\ \bibnamefont {Wu}},\ }\href@noop {} {\bibfield
  {journal} {\bibinfo  {journal} {Ann. Phys.}\ }\textbf {\bibinfo {volume}
  {327}},\ \bibinfo {pages} {2955} (\bibinfo {year} {2012})}\BibitemShut
  {NoStop}%
\bibitem [{\citenamefont {Griffiths}(1995)}]{griffiths}%
  \BibitemOpen
  \bibfield  {author} {\bibinfo {author} {\bibfnamefont {D.~J.}\ \bibnamefont
  {Griffiths}},\ }\href@noop {} {\emph {\bibinfo {title} {Introduction to
  Quantum Mechanics}}}\ (\bibinfo  {publisher} {Prentice Hall},\ \bibinfo
  {address} {Upper Side River, NJ},\ \bibinfo {year} {1995}),\ pp.\ \bibinfo
  {pages} {68--78}\BibitemShut {NoStop}%
\bibitem [{\citenamefont {Borowski}\ \emph {et~al.}(2003)\citenamefont
  {Borowski}, \citenamefont {Gemmer},\ and\ \citenamefont {Mahler}}]{pressure}%
  \BibitemOpen
  \bibfield  {author} {\bibinfo {author} {\bibfnamefont {P.}~\bibnamefont
  {Borowski}}, \bibinfo {author} {\bibfnamefont {J.}~\bibnamefont {Gemmer}}, \
  and\ \bibinfo {author} {\bibfnamefont {G.}~\bibnamefont {Mahler}},\
  }\href@noop {} {\bibfield  {journal} {\bibinfo  {journal} {Europhys. Lett.}\
  }\textbf
  {\bibinfo {volume} {62}},\ \bibinfo {pages} {629} (\bibinfo {year}
  {2003})}\BibitemShut {NoStop}%
\end{thebibliography}
%
\end{document}